\documentstyle[epsf]{myl-aa} %for a non-referee version
\begin{document}

   \thesaurus{
              (02.01.1;  % Acceleration of particles
               09.03.2;  % cosmic rays
               10.08.1   % Galaxy: halo
               11.10.1;  % Galaxies: jets
               13.07.1)} % Gamma rays: bursts
\title {Galactic $\gamma$-ray bursters - \\
      an alternative source of cosmic rays at all energies }

   \author{Arnon Dar\inst{1,2} and
           Rainer Plaga\inst{2}}

\offprints{plaga@mppmu.mpg.de}
  
\institute{ Department of Physics and Space Research Institute, Technion, 
Haifa 32000, Israel\and
Max-Planck-Institut f\"ur Physik, F\"ohringer Ring
6, D-80805 M\"unchen, Germany }

   \date{Received February 10 1999 / Accepted June 22 1999}
     
\maketitle
\markboth{A.Dar \& R.Plaga:
An alternative source of cosmic rays at all energies }
{}

\begin{abstract}
We propose a new hypothesis for the origin of 
the major part of non-solar
hadronic cosmic rays (CRs) at all energies: highly relativistic,
narrowly collimated jets from the birth or collapse of neutron
stars (NSs) in our Galaxy 
accelerate ambient disk and halo matter to CR energies
and disperse it in hot spots which they form when they stop in the
Galactic halo. Such events are seen as cosmological gamma-ray bursts
(GRBs) in other galaxies when their beamed radiation happens to point
towards Earth. 
This source of CRs is located in the Galactic halo.
It therefore explains the absence of the
Greisen-Zatsepin-Kuz'min cutoff in the spectrum of the 
ultra-high energy CRs.
The position in energy of the ``ankle'' in the CR energy spectrum
is shown to arise in a natural way.
Moreover, an origin of lower energy
CRs in the Galactic halo naturally accounts
for the high degree of isotropy of CRs
around 100 TeV from airshower observations,
and the small galactocentric gradient
of low-energy CRs derived from gamma-ray observations. 
\end{abstract}
%
%________________________________________________________________ 
\section{Introduction}
The origin of non-solar cosmic rays, discovered in 1912, is a
major unsolved puzzle in physics. Recent observations
do not confirm some expectations 
in the currently accepted framework for hadronic CR origin,
where cosmic-ray nuclei with energies
below $3\times 10^{15}$ eV (the ``knee'') are accelerated in Galactic
supernova remnants (SNRs) (\cite{ginz}), and those above $3\times 10^{18}$
eV (the ``ankle'') (UHE CRs),
for which a disk origin is unlikely due to their
isotropy, in sources far beyond our Galaxy (\cite{burb}). 
Some observational anomalies are:
\begin{itemize}
\item The ``GZK cutoff'' in the CR
intensity at energies above $\sim 10^{20}$ eV, 
due to interactions of the extragalactic CRs with the
cosmic microwave radiation, 
as predicted by Greisen (\cite{greisen}) and by
Zatsepin \& Kuz'min (\cite{zatsepin}) 
seems to be absent (\cite{agasa1}). At the same 
%The plasmoids ejected by this object do remain
%confined over a distance of about 0.1 Ly with
%no sign of disintegration.
time the directions of the events with the highest energy
do not line up with nearby potential
extragalactic CR sources. It is difficult to understand 
these two observations within any model of an extragalactical
origin of UHE CRs (\cite{Hillas}).
\item No very-high energy $\gamma$-rays, 
due to proton interactions, are observed
from some SNRs, for which
a measurable flux is expected in simple
models (\cite{drury2}) within the conventional framework 
(\cite{gam1}; \cite{gam2}; \cite{gam3}).
\item Detailed models (e.g., 
(\cite{galradius})) 
\newline
yield a significantly 
larger galactocentric gradient 
in the sky distribution of high-energy ($>$100 MeV) $\gamma$-rays 
from interaction of CRs from SNRs in the Galactic interstellar 
medium than the one observed by the EGRET detector
on the Compton Gamma-Ray Observatory
(\cite{hunter}; \cite{strong2}; \cite{erlykin2}).
\item Diffusive propagation of CRs
from sources sharply concentrated towards the Galactic disk 
and centre (such as SNRs)
yields anisotropies in the distribution of charged CRs
at an energy of about 
100 TeV in excess of the observed value (\cite{aglietta}) by more than an
order of magnitude. This is 
consistently found both in simple order-of-magnitude
estimates (\cite{snowm}) and sophisticated calculations of CR propagation
(e.g., \cite{voelkpropa}). 
\end{itemize}
The last three observations suggest that, perhaps, SNRs 
are not the main accelerators of Galactic hadronic CRs at energies below
the knee.
\\
Maybe these anomalies can be resolved
in more sophisticated models 
within the conventional framework; e.g. 
by modifying standard ideas about 
intergalactic CR propagation for the first (\cite{sigl})
and interstellar CR propagation
for the third and fourth (\cite{wolfendale}) anomaly, 
respectively. Spectral cutoffs at energies much below the knee
could be the reason
for the absence of TeV $\gamma$-rays from 
some fraction of SNRs (\cite{baring0}).
Here we explore the
possibility that the anomalies are first indications that
the current framework for the origin of hadronic CRs is wrong,
and propose an alternative origin. 
\\
Gamma-ray bursts (GRBs) are short and intense bursts of $\gamma$-rays
from distant galaxies that are detected by space satellites at a rate of
$R_{\rm GRB}$ $\sim 10^3$ per year (\cite{grb}).  
Their origin is unknown, but
various observations suggest an association with the formation of
black holes or the birth or collapse 
of neutron stars (NSs) due to mass accretion or phase transition, in close
binaries or alone.
The strongest indication for a connection with the birth
of NSs are the coincidences of supernovae with GRBs, e.g.
 SN 1998bw and GRB 980425 (\cite{galama2})
and of SN 1999E and GRB 980919 (\cite{thorsett};
\cite{kulkarni2}).
A connection between events in the life of NSs  and 
GRBs at cosmological distances was first suggested
by Paczynski (\cite{pac}), Goodman et al. (\cite{good}) and
Dar et al. (1992).
\\
The rest of the introduction discusses the assumptions we 
make about GRBs in our scenario, beaming in GRBs and 
previous work on CR acceleration
in GRBs. Sect. 2 turns to a more detailed quantitative discussion
of our picture of GRBs. Sect. 3 discusses the propagation
of the ejecta of GRBs in the Galaxy and Sect. 4 treats
energy spectrum of CRs which is expected in our scenario.
Sect. 5 and 6 describe how our
scenario resolves the above-mentioned anomalies of UHE CRs and lower energy
CRs respectively. Sect. 7 concludes the paper.

\subsection{Basic assumptions for our scenario}
We make two basic assumptions for our scenario.
\begin{itemize}
\item GRBs are not rare events, but common phenomena
in the life of neutron stars; i.e., 
in our Galaxy they occur with a frequency
comparable to the NS birth rate.
We argue in the next subsection that a 
necessary implication from this and the observed
rate of GRBs $R_{\rm GRB}$
- that the emission from
GRBs is strongly beamed - seems likely in view
of recent GRB observations.
\vskip 0.2in
\item Jetted ultra-relativistic
ejecta (``plasmoids'') emit
the beamed radiation in GRBs and carry a considerable
fraction of the total binding energy of a NS. The
ejecta remain confined to a small size until slowed
down - via interaction with an ambient medium - to mildly
relativistic speeds. Only then do they dissipate most of
their kinetic energy - partly by particle acceleration - 
and release most of the previously accelerated particles. 
We give arguments in favour
of this assumption - based on
a special interpretation of NS kicks and
analogies with similar astrophysical systems -
and describe it in more detail in Sect. \ref{quantprop}. 
\end{itemize}
We return to the question whether these assumptions
are overly speculative in the conclusion.
\subsection{Beaming in GRBs}
If the true rate of GRBs is similar to the birth 
rate of NSs, $R_{NS}\simeq 0.02~{\rm year}^{-1}$ in 
galaxies like our own (\cite{bergh}), the radiation emitted in 
GRBs must be narrowly beamed into  
a solid angle $\Delta\Omega\simeq\pi\times 10^{-6}$ in order 
to explain their rate inferred from $R_{\rm GRB}$ (\cite{wijers}) 
in such galaxies: 
\begin{equation} 
R_{\rm GRB}({\rm MW})\simeq
2R_{\rm NS}(\Delta\Omega/4\pi)\sim 10^{-8}~{\rm year}^{-1}.
\end{equation} 
Indeed GRBs could be
produced by highly relativistic, 
collimated
jets with bulk motion Lorentz factors
$\Gamma=1/\sqrt{1-v^2/c^2}\sim 10^3$
(\cite{shaviv95}; \cite{blackman96}; 
\cite{reesjet96}; \cite{chiang}; \cite{arnon98}). 
\par\noindent
In this connection - and always in this paper - the term
``jet'' is meant to imply only collimated ejecta into a small
opening angle 
%(arbitrarily defined as $<$ 15$^{\circ}$ according 
%to Bridle $\&$ Perley (\cite{bridle}))
and not an emission stationary in time.
The very large redshifts recently measured in some
GRBs and their host galaxies  
- e.g. z=3.42 for GRB 971214 (\cite{kulkarni98}) - seem firmly to 
require jetted emission 
for energetic reasons.
%; \cite{fruchter98a}; \cite{fruchter98b}; 
%\cite{djorgovski98}; \cite{bloom98})
%suggest that GRBs are jetted (\cite{arnon98}). 
For instance, the assumption of isotropic emission implies  
that $\ge 4\times 10^{54}$ erg - which is 
hardly obtainable in the birth
or collapse of any stellar object - 
was emitted by GRB 990123
as $\gamma$-rays (\cite{djorgovski99}; \cite{kippen99}),
while only
$\sim 10^{48}$ erg was emitted if the emission
was narrowly beamed into a solid angle $\Delta\Omega\sim\pi\times 
10^{-6}$. Hjorth et al.(\cite{hjorth}) found a complete absence
of optical polarisation in GRB 990123 and mention
a relativistic jet which is seen at a small viewing angle
as one possible explanation for their somewhat surprising
observation.
Emission from ``narrow'' jets
with bulk motion Lorentz factors $\Gamma\sim 10^{3}$
is indeed beamed into a solid angle $\Delta\Omega\sim
\pi/\Gamma^2\simeq\pi\times 10^{-6} $ consistent
\footnote{Such
Lorentz factor values are required from GRB observations
independent of whether beaming takes place (see, e.g.,  \cite{baring}).}
with Eq.(1). 
Here ``narrow'' means that the angle of the emitted ejecta
$\theta$ subtended from the burst site is smaller than 1/$\Gamma$.
There is some evidence that under the assumption of
jetted ejection this condition is fulfilled in GRBs (though
the following cannot be considered as proof for jetted ejection):
Schaefer \& Walker (\cite{schaefer}) argued from timing data
of the BATSE detector that for the burst GRB 920229, $\theta$ $<$
1$'$; this fulfils the ``narrow''-jet condition for
$\Gamma$ $<$ 3000.
The amount of beaming and physical structure of the
GRB jet assumed in our paper is very similar 
to the one in a model of Chiang \& Dermer (1997) for GRB afterglows.

If such strong beaming is typical - and this we assume - 
we observe only a very small fraction of a 
large rate of the events that can produce GRBs. We call these events
``Galactic'' 
(if they occur in our Milky Way (MW) galaxy) and ``Cosmological'' 
(if they occur in distant galaxies) gamma-ray bursts'' (GGRBs and CGRBs),
respectively.

\subsection{CR acceleration in GRB ejecta}
Gamma-ray bursters
have been proposed as efficient accelerators of ambient material to CR
energies before, but
GGRBs with isotropic CR emission
(\cite{grb1}; \cite{grb4}) 
cannot supply 
the bulk of the Galactic cosmic rays for energetic reasons.
Nor can the idea of UHE CRs from
CGRBs (\cite{grb2}; \cite{grb3}; \cite{boettcher}) explain
the absence of the GZK cutoff (\cite{grbeg}). 
In this paper we show that 
the effects of strong beaming make GGRBs 
plausible sources of non-solar hadronic cosmic rays
at {\it all} energies
(from $\simeq$ 100 MeV to 3 $\times$ 10$^{20}$ eV) observed near Earth. 
\begin{figure}[ht]
\vspace{0cm}
\hspace{0cm} \epsfxsize=9.1cm 
\epsfbox{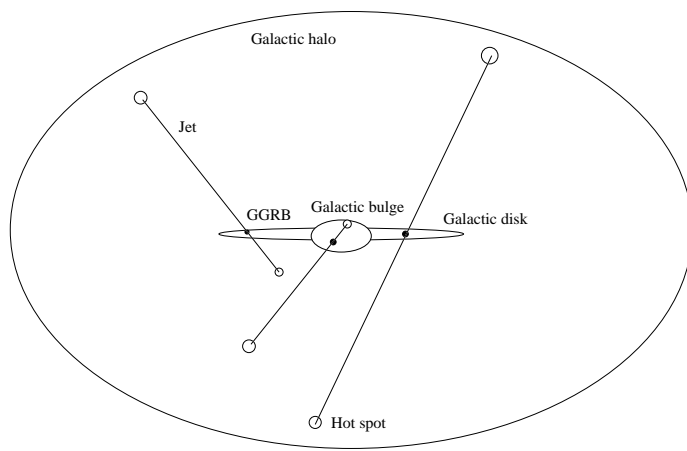}
\vspace{0cm}
\caption{ A highly schematic sketch of our scenario. The
birth or collapse of NSs in the disk of our Galaxy leads to an
ejection of two opposite jets that produce ``hot spots'' when they stop in
an extended Galactic halo.
}
\label{fig1}
\end{figure} 

\section{Quantitative properties of the plasmoids ejected in GRBs}
\label{quantprop}
The ejection of highly relativistic jets from accreting or collapsing
compact objects, and especially the physical mechanism for
their observed collimation ``remains a major unsolved
problem in astrophysics (\cite{longair})''.
Therefore, instead of relying on
theoretical models, we have tried to estimate the properties of 
those jets that may
produce the GRBs and cosmic rays from data
about NSs and
- by way of analogy -
from observations of similar jets.
\par\noindent 
NSs are observed to possess a large mean
velocity of $v\approx 450\pm 90~ {\rm km~s^{-1}}$ (\cite{nskick}).
Burrows \& Hayes (\cite{burrows}) argue that an anisotropic 
stellar collapse - rather than proper motions
due to orbital speeds in binaries - is the mechanism responsible
for the necessary ``kick'' and
and that matter - rather than neutrinos - carries most of the momentum.
Janka \& Raffelt (\cite{janka}) find that
works in which deformed
neutrino spheres impart large kicks to NSs
are incorrect and that it is in general difficult
to impart the necessary momentum via neutrinos.
In this light - and making the above assumptions - 
it seems plausible that
the ejection of relativistic jets in the birth or collapse of NSs is
responsible for their observed large mean velocity. 
Momentum conservation then implies that
the kinetic energy of the jets is 
\begin{equation}
E_K\approx cP\sim vM_{NS}c\sim 4\times10^{51}~{\rm erg},
\label{jetenergy} 
\end{equation}
where we used the typical observed NS mass, $M_{NS}=1.4M_\odot$. If two
antiparallel jets are ejected, then the above estimate becomes a lower
limit for the jet kinetic energy, that may be of the order of $E_{\rm k}\sim
10^{52}~{\rm erg}$.  Thus, already a modest 
fraction $f\simeq 0.01$ of the total energy
injected into the MW in jets from GGRBs, 
at a rate $R\sim 0.02~{\rm year}^{-1}$,
similar to the NS birth rate, can supply the estimated
Galactic CR luminosity 
$L_{MW}[CR]\sim 1.5\times
10^{41}~{\rm erg}$ (e.g., \cite{drurv}). 
\par\noindent 
Highly relativistic jets are emitted by all astrophysical systems
where mass is accreted at a high rate from a disk onto a central black
hole (BH). They are observed in Galactic superluminal sources, such as the
micro-quasar GRS 1915+105 (\cite{mirabelnat}), where mass is accreted onto a
stellar BH, and in many active galactic nuclei (AGN), where mass is
accreted onto a super-massive BH (\cite{cyga}). 
Mildly relativistic jets from mass
accretion were seen both in AGN and in star binaries containing NSs such
as SS433 (\cite{margon}) and Sco X-1 (\cite{fomalont}). 
\\
High-resolution radio observations of the
micro quasar GRS 1915+105 resolved the
narrowly collimated relativistic jets into clouds of plasma (plasmoids)
that are emitted in injection episodes which are correlated with sudden
removal of the accretion disk material (\cite{mirabelnat}). 
In GRS 1915+105, during the first five
days after ejection, these plasmoids
appear to have expanded with the speed
of sound in a relativistic gas, $c_{\rm s}\sim c/\sqrt{3}$ 
in the plasmoid rest frame, 
corresponding to
vertical expansion speed of $\sim c/\Gamma\sqrt{3}$ in the lab frame), until
they reach a radius
$R_{\rm p}\sim 2\times 10^{15}~{\rm cm}$ ($\sim 0.001~{\rm pc}$)
(\cite{mirabel2}; \cite{mirabelnat}).
Their expansion seems to be slowed afterwards, probably 
due to some confinement mechanism (\cite{mirabelnat}; \cite{mirabel3}). 
Likewise, jets from quasars and radio galaxies can also often be resolved
into the ejection of distinct clouds at the pc scale 
(e.g., \cite{cloud})
and might be governed by mechanisms similar 
to the jets emitted by micro quasars. 
These jets retain a
constant radius after initial expansion
(see, e.g., (\cite{swain})) and some confinement mechanism is
clearly at work.
The repeated ejection of plasmoids (often called ``plasmons'' in the
older literature) is a well known model
for extragalactic jets (\cite{cyga}; \cite{deyoung}; \cite{chris}).
\\
The confinement mechanism which collimates plasmoids
is not understood; two possibilities are magnetic
confinement (\cite{cyga}) and inertial confinement via the ram pressure
of the ambient medium (\cite{deyoung}). 
The plasmoids slow down by sweeping up the
ionised interstellar medium in front of them. 
\\
It is one of our assumptions
that the plasmoids from the birth/collapse of NSs evolve in
a similar way to the ones apparently ejected in galactic superluminal
sources and AGNs (\cite{arnon98}). 
The plausibility of this idea
was recently endorsed by
Mirabel \& Rodriguez (1999) 
in their review about relativistic
jets the Galaxy; they write:
``the study of the less extreme collimated outflows
in our own Galaxy may provide clues for a better understanding
of the super-relativistic jets associated with ... GRBs''.
The plasmoids in GRS1915+105 could only be followed up
to distances of about 0.03 pc from the central
source. 
A non-thermal jet with a
radius of about 0.1 pc at a distance of about
60 pc from the central source GRS1915+105
and pointing back to it was observed recently in the radio
range (\cite{mirabel4}).
No connection between the central
source and this jet was observed.
A possible interpretation is that this jet is a collimated
plasmoid that was ejected by the central source.
These various observations of jets motivate the first 
part of our second assumption in the Introduction that:
\begin{itemize}
\item after an initial expansion
- and until their final dissipation phase - 
the plasmoids retain a radius of
about 0.01 pc, collimated by the as yet ill-understood
mechanism apparently at work in the plasmoids and jets discussed above.
\end{itemize}  
Plasmoids ejected in the birth/collapse of NSs
- as opposed to the ones in micro quasars and AGNs -
are ``isolated'', i.e. their central source is active on a
time scale smaller then their propagation time.
Our assumption is that this fact
does not invalidate the qualitative similarity between
the mentioned three object classes.
\\
In extragalactic double-lobe radio galaxies like Cyg-A (\cite{cyga})
it is observed that apparently bulk kinetic  
energy of the jet is dissipated mainly in a ``hot spot''.
The nonthermal jet at a large distance
from GRS1915+105, mentioned above, is an indication
that ``isolated'' plasmoids might show a similar behaviour.
A rising internal kinetic
pressure - due to the sweep up of relativistic particles -
and dissipation of their internal magnetic energy 
might lead finally to
their expansion and stopping in ``hot spots'', where 
a considerable part of their initial
kinetic energy is released in highly relativistic particles.
Another factor leading to a final expansion
could be a decrease in inertial confinement
due to the a falling ram pressure in decelerating plasmoids, but
the detailed physics of hot spots produced by ``isolated'' plasmoids
remains unclear.
This set of observations motivates the second part 
of the second assumption in the Introduction:
\begin{itemize}
\item the plasmoids dissipate most of their energy 
in turbulent motion of inter-halo
matter and CR production only when they have been slowed
down - via sweeping up ambient matter - to mildly relativistic energies. 
Most of previously accelerated particles are also released in the phase. 
\end{itemize}
In the case of a GGRB there is no stationary flux of particles 
whose energy is dissipated
at a relatively sharply defined ``working surface'', as in the case
of the jets of active galaxies (\cite{cyga}).
Under the above assumptions most of the dissipation and
particle acceleration will
take place in a region which is small compared with
the total distance travelled by the plasmoid and which we 
might still call ``hot spot''.
\\
The possibility that relativistic jets responsible for GRBs 
are sometimes formed in
supernova (SN) explosions has been discussed by 
various authors (\cite{cen}; \cite{wang}; \cite{eichler}).
It is conceivable that a considerable fraction of SN explosions
leads to GRBs with the properties required in our scenario.
The jet like features observed in many SNRs
(\cite{gaensler}) 
may even hint towards this possibility (though they certainly do
not prove it).
However, this idea is not mandatory for our scenario; other events in
the life of NSs might be responsible for most GRBs.
\\
Because the accretion rates and magnetic fields involved in
events leading to GRBs
are larger than in all other systems discussed above, an initial
Lorentz factor of $\Gamma\sim 10^3$, as required by GRBs observations, 
does not appear excessive.
The initial equipartition magnetic fields in such plasmoids
can be estimated to be $B> 100~{\rm G}$.
Extragalactic jets, which might be qualitatively 
similar objects, have been identified as prolific CR
accelerators (\cite{burb}), in particular their hot spots (\cite{bira}).
Highly magnetised relativistic
plasmoids are both efficient CR accelerators (\cite{chiang}; \cite{grb5})
(through
Fermi acceleration) and strong emitters of beamed $\gamma$-rays through
synchrotron emission, inverse Compton scattering and resonance scattering
of interstellar light. When they point in (or precess into) our direction
from external galaxies, they produce the observed GRBs and their
afterglows.
\section{Propagation of the ejected plasmoids in the Galaxy}
For our discussion, we adopt an extremely simplified picture of our
Galaxy, namely, a Galactic thin disk surrounded by a spherical halo with
an approximate radius of $R_{\rm h}\sim 50~{\rm kpc}$ 
( $V_{\rm h}\sim 1.5\times
10^{70}~cm^3$) and a mean gas density of $n_{\rm h}\sim 10^{-3}~{\rm cm}^{-3}$.
The highly relativistic plasmoids which are emitted with $E_k\sim
10^{52}~{\rm erg}$ perpendicular to the Galactic disk are 
assumed to be decelerated mainly by sweeping up ambient matter.
Whether this is an accurate description depends on whether
the swept-up material ``sticks'' to the plasmoid and is
incorporated into it, rather than e.g. forming a hot expanding wake
behind the plasmoid. A full dynamic treatment of the propagation
of ultrarelativistic plasmoids is required, a problem that is however
beyond the scope of this paper. 
The plasmoids then stop in the Galactic
halo, when the rest mass energy of the swept-up ambient material is equal
to their initial kinetic energy:  The column density of gas required to
stop jets with a radius smaller than $R_{\rm p}\sim 0.01~{\rm pc}$ is 
\begin{equation}
N_{\rm c}\geq 2.2[(E_{\rm k}/10^{52}~{\rm erg})/(R_{\rm p}/0.01~{\rm pc})^2)]
\times 10^{21}~{\rm cm}^{-2}. 
\label{sweep}
\end{equation}
The mean column density of gas (mainly molecular, atomic and ionised
hydrogen) perpendicular to the Galactic disk, is
$N_c<10^{21}~{\rm cm}^{-2}$ (\cite{bere}). 
From Eq.(\ref{sweep}) one reads that
the jets stop in the Galactic halo when
their radius reaches $R_p\sim 0.1 {\rm pc}$ 
(if $n_{\rm h}\geq 10^{-4}~{\rm cm}^{-3}$) 
and form hot spots. This is illustrated in Fig 1.

The ``decelerating plasmoid
model'' of Chiang \& Dermer (1997) for GRB afterglows
makes similar assumptions and works out
some consequences in more numerical detail. 
In particular, in their model the
plasmoid expands only slowly, and decelerates
mainly via sweeping up ambient matter.
In their case a plasmoid
with a radius of about 10$^{-4}$ to 10$^{-3}$ pc,
ejected by the CGRB with a kinetic energy of 10$^{50}$ ergs,
travels 30 - 100 pc in the first
3 - 10 days after the GRB (measured in the observer frame)
and does not slow down
to non-relativistic speeds during the simulated propagation.

\section{Energy spectrum of CR produced by the plasmoids}
\subsection{High-energy limit for the acceleration of CR in the plasmoids}
The typical
equipartition magnetic fields in the hot spots 
discussed in the previous section may reach $B\sim
(3E_{\rm k}/R_{\rm p}^3)^{1/2}\sim 1~{\rm G}$. Synchrotron losses cut off
Fermi acceleration
of CR nuclei with mass number A at  $E\sim \Gamma
A^2Z^{-3/2}(B/G)^{-1/2}\times 10^{20}$ eV.  Particle escape cuts off Fermi
acceleration when the Larmor radius of the accelerated particles in the
plasmoid rest frame becomes comparable to the radius of the plasmoid,
i.e., above  $ E\simeq \Gamma Z(B/G)(R_p/0.1~pc)\times 10^{20}$~
eV. In the hot spots, $R_p\sim 0.1$ pc, $\Gamma\sim 1$ and $B\sim 1~ $G.
Consequently, CR with $E>Z\times 10^{20}$~eV can no longer be isotropised
by acceleration or deflection in the hot spots. Thus, CR with energies
above $10^{20}~$eV are heavy nuclei in our scenario.

\subsection{The energy spectrum at lower energies}
\label{seclowe}   
Fermi acceleration in or by the highly relativistic jets from GRBs could
produce a broken power-law spectrum, $dn/dE\sim E^{-\beta}$, with
$\beta\sim 2.2$ below a knee around $E_{\rm knee}\sim A~{\rm PeV}$ 
and $\beta\sim 2.5$ 
above this energy (\cite{grb5}). Spectral indices $\beta$ $\sim$ 2-3 are
expected in relativistic shock acceleration with tangled magnetic
fields (\cite{ballard}). Galactic magnetic confinement increases the density
of Galactic CR by the ``concentration factor'' c = $c\tau_{\rm h}/R_G$, 
where $\tau_{\rm h}(E)$ is the mean
residence time in the halo of Galactic CR with energy $E$ and $R_G$ is the
radius of the Galactic magnetic-confinement region. With the standard
choice for the energy dependence of the 
diffusion constant (observed, e.g.,
in solar-system plasmas (\cite{bere})) one gets: $\tau_{\rm h}\propto
(E/Z)^{-0.5}$.  Consequently, the energy spectrum of CR is expected to be
\begin{equation} 
dn/dE\sim C (E/E_{\rm knee})^{-\alpha} 
\end{equation} 
with $\alpha\simeq \beta+0.5\simeq 2.7~ (\simeq 3)$ below 
(above) the knee. The normalisation constant cannot yet
be determined - except by the rough
total-energy argument mentioned 
in Sect. \ref{quantprop} after Eq.(\ref{jetenergy}) -
but is universal for all energies.
In this respect our scenario and the
conventional framework are in the same status.
  
This power-law is expected to continue as
long as the Galactic magnetic field confines the CRs, i.e.,
up to energies near the ``ankle''. 
\begin{figure}[ht]
\vspace{0cm}
\hspace{0cm}\epsfxsize=10.1cm \epsfbox{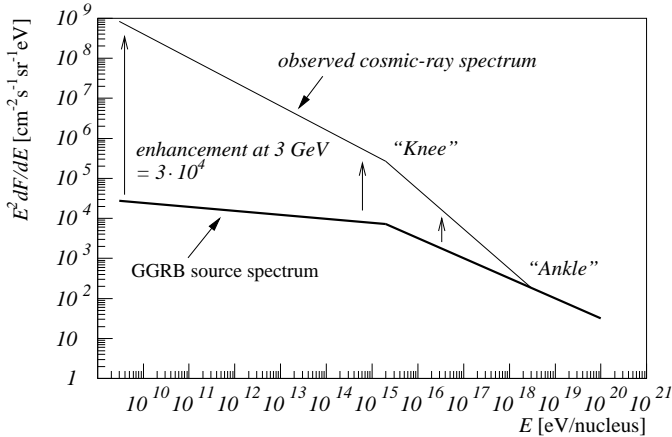}
\vspace{0cm}
\caption{ The observed flux of cosmic rays
(thin line) as a function of primary energy E 
is well described by a power law that changes
its slope sharply at only two
energies, the ``knee'' and the ``ankle''.
At energies below the ankle
it is enhanced (by a factor (E/E$_{ankle}$)$^{-0.5}$)
over the GGRB source spectrum
(thick line, a power law with differential power law index
of -2.2 below the knee and $\simeq$ -2.5 above it)
by way of trapping in the
Galactic halo magnetic fields}
\label{fig2}
\end{figure}
\noindent 
\subsection{Explanation
of the ``ankle''}
Part of the kinetic energy released by GGRBs is transported into the Galactic
halo by the jets. Assuming equipartition of this energy, without large
losses, between CR, gas and magnetic fields in the halo during the
residence time of CR there, the 
magnetic field strength  B$_{\rm h}$ in the halo is
expected to be comparable to that of the disk $B_{\rm h} \sim
(2L_{\rm MW}[CR]\tau_{\rm h} /R_{\rm h}^3)^{1/2}\simeq 3~\mu G$
 where $\tau_{\rm h}\sim 5\times
10^9~{\rm years}$ is the mean residence 
time of the bulk of the CR in the Galactic halo
(see Sect. \ref{kneesec})
cosmic rays with Larmor radius larger than 
the coherence length $\lambda$
of the halo magnetic fields (\cite{clay}), i.e., with
energy above  
\begin{equation} 
E_{\rm ankle}\sim 3\times 10^{18}
(ZB_{\rm h}/3\mu{\rm G})(\lambda/{\rm kpc})~{\rm eV},
\label{ankle} 
\end{equation}
escape Galactic trapping.
Our ``preferred'' values indicated in Eq.(\ref{ankle}) are
well within the range of possible values preferred by other
researchers (\cite{clay}).
Thus, the CR ankle is explained as the energy where the mean residence time
$\tau_{\rm h}$(E) of CR becomes 
comparable to the free escape time from the halo
$\tau_{\rm free}\sim 1.6(R_{\rm h}/50~{\rm kpc})\times 
10^5~{\rm years}$.
This explanation for the ankle was already proposed by
one of us (\cite{plaga}) 
in the context of a proposed extragalactic
origin of hadronic CRs, to which the present
proposal bears of course some similarity (in both cases all 
hadronic CRs have a common origin outside the Galactic disk).
\\
The spectrum of 
CRs with energies above the ankle that do not suffer Galactic
magnetic trapping is then the CR spectrum produced by the jet, i.e.,
\begin{eqnarray}
dn/dE\sim C 
(E_{\rm ankle}/E_{\rm knee})^{-3}(E/E_{\rm ankle})^{-2.5};
\nonumber
\\
~~~~ E> 
E_{\rm ankle}.
\end{eqnarray}
This predicted spectral index of 2.5 has to be compared
with the one 
recently determined by the AGASA collaboration (\cite{agasa1})
of 2.78$^{+0.25}_{-0.33}$.
In the conventional framework - where
only UHECRs have an extragalactic origin - the position
of the ankle at just the energy where Galactic magnetic fields
can no longer confine CRs is pure coincidence.
\subsection{The ``knee'' in the CR spectrum and low-energy 
propagation}
\label{kneesec}
The knee in the CR source spectrum still lacks a complete explanation (see
however Dar (1998b)), but we think that a single source population
over the whole CR energy range offers a more natural framework to
understand the CR spectrum than the prevailing one where a ``fine tuning''
in the intensity of different source components is often required at the
knee. The ``concentration factor'' c (see Sect. \ref{seclowe} above)
over the GGRB source spectrum reaches
$3\times 10^{4}$ at GeV energies (see Fig. 2),
corresponding to a confinement time in
the Galactic halo of about $5\times 10^{9}~{\rm years}$. We interpret
the experimentally measured ``lifetime'' of low-energy CR, of the order
$2\times 10^{7}~{\rm years}$ (\cite{bere}; \cite{connel}), as the residence
time in the Galactic disk, in which we are located as observers, rather
than the time since acceleration, as in most other models.  The mean
residence time $\tau_{\rm d}$ of low-energy particles in the disk is given as: 
\begin{equation}
\tau_{\rm d} \simeq \tau_{\rm h} \cdot 
(d_{\rm d}/d_{\rm h})^2 (D_{\rm h}/D_{\rm d}).
\end{equation}
If the diffusion coefficient $D_{\rm d}$ in the disk is smaller than that in
the halo ($D_{\rm h}$) by about a factor 10$^3$ - 
due to a stronger turbulence in
the disk induced by SN explosions - 
then $\tau_{\rm d}$ is of the observed order
of magnitude.~

\section{A prediction of our scenario:
angular clustering of UHE CRs}
Small deflections of UHECR by turbulent magnetic fields along their
arrival trajectory, with a length $d$, spread their arrival times over
\begin{eqnarray}     
\Delta t\sim 750(d/50~{\rm kpc})^2(\lambda/{\rm kpc}) \times 
\nonumber
\\
(ZB/3\mu
{\rm G})^2(E/100~{\rm EeV})^{-2}~{\rm years}.  
\end{eqnarray} 
The nearly isotropic release of CR in the halo by GGRBs and the spread in
their arrival times, during which many GGRBs occur, accounts for the
nearly isotropic sky distribution of the UHECR.
Jets which stop in the
Galactic bulge or the Galactic disk (see Fig. 1) may produce a detectable
enhancement in the flux of UHECR from these directions. 
\\ 
First data from the Akeno Giant Air Shower Array (AGASA) (\cite{agasa2})
suggested a clustering in the arrival directions of the UHECR, these
indications have recently been reaffirmed with an enlarged database
(\cite{agasa3}). 
The directions of these clusters 
do not coincide with any known nearby extragalactical objects
which would seem a plausible candidate for
acceleration the UHECR; this is consistent with a 
Galactic halo origin. 
%s Such
%clustering is not possible if the origin of the UHECR are particles from
%the decay or annihilation of relic particles\cite{Hillas} with mass 
%smaller than the
%Planck mass, $M_{\rm p}=
%\sqrt{\hbar c/G}\approx 1.22\times 10^{19}~{\rm GeV}/c^2$,
%which are distributed isotropically in the sky: Even if all the rest mass
%energy was converted into two opposite jets of $10^{19}~{\rm eV}$ protons with
%transverse momentum of $\sim 1~{\rm GeV}/c$ with respect to the jet axis, the
%mean number of UHE CR events that such a jet from a typical halo distance
%of $30~{\rm kpc}$ could produce 
%in the AGASA detector (100 ${\rm km}^2$) is still only
%$\sim 3\times 10^{-6}$. 
Such clustering is predicted if the UHE CRs
are produced/released in hot spots in the Galactic halo. Their accumulated
r.m.s. deviation angle over a distance $d$ by random walk in the turbulent
magnetic field is
\begin{eqnarray}  
\theta\sim 5.4^0(d/50~{\rm kpc})^{1/2}(\lambda/{\rm kpc})^{1/2}(ZB/3\mu
{\rm G}) \times
\nonumber
\\
(E/100~{\rm EeV})^{-1}.  
\end{eqnarray} 
Therefore, if the origin of UHECR is $\sim 0.1~{\rm pc}$-sized hot spots in the
halo, the arrival direction of CR with energies around $100~ {\rm EeV}$ is
predicted to cluster around $\sim$ 30 (arrival time spread $\times$ GGRB rate)
fixed hot spot positions in the sky
whose number and spread around them decrease with increasing energy.
\\
An origin of UHE CRs from the decay or annihilation of ultra-massive
new particles in the Galactic halo has recently been discussed 
(\cite{Hillas}; \cite{bere2}; \cite{kuzmin}).
In such a scenario clustering 
is not possible if the the UHE CRs are nucleons from the decay or
annihilation of relic particles with mass smaller than the Planck mass,
$m_{\rm p}=\sqrt{\hbar {\rm c/G}}\approx 1.22\times 10^{19}~
{\rm GeV/c}^2$, which are
distributed isotropically in the sky. Even if all the rest mass energy was
converted into two opposite jets of 
$10^{19}~{\rm eV}$ nucleons with transverse
momentum of $\sim 1~GeV/c$ with respect to the jet axis, the mean
number of UHE CR events that such a jet from a typical halo distance of
30 kpc could produce in the AGASA detector (100 km$^2$) is still
only $\sim 5\times 10^{-6}$.
 
\section{Some experimental data on CRs compared to 
expectations in our scenario}
The CR spectrum which is suggested in our scenario is displayed
in Fig. 2. 
As discussed in the previous sections 
it agrees well with the measured spectrum. There is no GZK cutoff
because of the small distance to the sources
(O(50 kpc)), over which absorption
of UHE nucleons in the microwave-background radiation
is negligible.
As the major part of the hadronic CRs are not produced in SNRs
in our scenario, the predicted flux of high-energy $\gamma$-rays
due to proton interaction from these objects can be low,
thus resolving the second anomaly mentioned in the introduction.
%%Neglecting fragmentation
%%and attenuation due to collisions with background photons and gas
%%particles, which become important above $10^{17.5}$~eV, the abundances of
%%CR nuclei is enhanced by a factor $Z^{0.5}$ compared to that of swept up
%%ISM, due to Galactic magnetic trapping. 
%(detailed calculations of the
%chemical composition of CR are beyond the scope of this letter).
A  quantitative 
calculation of the expected density of hadronic CRs as a function 
of galactocentric radius in our scenario
is difficult
because of the ill-understood propagation properties
of the plasmoids in different part of the Galaxy (halo, bulge, etc.).
In first approximation this density is
predicted to fall with a halo scale (i.e., tens of kpc).
This is  in better agreement with
observations than the $\sim 5~ {\rm kpc}$ scale height
of SNRs, which is expected to be close to the characteristic
decay scale
in the conventional scenario (third
anomaly in the Introduction; \cite{galradius}\footnote{
The propagation models assuming a SNR origin of CRs
in this publication predict
decay scales between 4.5 to 9 kpc.}). 
Recent determinations of the gradient 
from EGRET data yield for the
exponential decay scale between 5 kpc and 20 kpc (read
off diagrams in these publications):
16 kpc (\cite{hunter}), 34 kpc (\cite{erlykin2}), 23 kpc (\cite{strong2}).
\\
Moreover 
- due to the near isotropy of their sources - the large-scale
anisotropy of CRs must be very small 
in our scenario at all energies.
This resolves the fourth anomaly mentioned in the Introduction. 

\section{Properties of decaying plasmoids in the halo}
Electrons suffer large energy losses by synchrotron emission and inverse
Compton scattering while diffusing out of their jet acceleration sites.
Thus, CR electrons in the Galactic disk could have their origin mainly in
Galactic SNRs (\cite{ginz});  this theory has recently received
experimental support from X-ray (\cite{allen}) and TeV $\gamma$-ray 
(\cite{tanimori}) observations of Cas-A and the remnant of SN 1006,
respectively. 
%In the hot spots - under conditions where a fraction 
%$\geq$ 0.01 of all
%ambient particles is accelerated in a relativistic
%shock - protons are probably accelerated with
%an efficiency a factor ``(proton mass/electron mass)'' larger than
%electrons. 
Totani (\cite{totani}) argued that synchrotron
radiation from accelerated protons is an important
emission mechanism in GRB afterglows.
Assuming hot spots accelerate
mainly baryons on a time scale
of about 1000 years over which they irradiate 10$^{50}$ ergs
in proton synchrotron radiation, their 
resulting radio intensity will be less than a mJy (\cite{plaga2}). 
The hot spots are thus not
expected to be particularly conspicuous radio sources. The
detailed observational
consequences of hot-spot remnants in the halo of the Milky Way
will be discussed elsewhere (\cite{plaga2}).

\section{Conclusion}
We proposed an alternative source for the origin
of hadronic CRs at all energies which is in better
accord with some observational facts 
than the conventional one.
Some of the assumptions 
on which it is based are certainly speculative.
However, our ideas seem conservative when compared to
the audacity of some proposals
by outstanding scholars to explain the puzzling
properties of ultra-high energy CRs - like, e.g., a violation
of Lorentz invariance (\cite{glashow}), 
the existence of a new class
of super-massive elementary particles 
in the Galactic halo (\cite{bere2}; \cite{kuzmin};
\cite{Hillas})
or supersymmetric light strongly-interacting particles (\cite{bierm2}).
Further observations of GRBs and their afterglows
will reveal if the contention that a GRB is
a common phenomenon
in the life of a neutron star 
- crucial for our scenario - is tenable.
The next generation of UHE CR detectors (\cite{mantsch}) 
will rule out or confirm
our prediction about their angular clustering properties.
Basic issues of our scenario which require more work
are the exact nature of the event giving rise
to a GRB and the observational
properties of ``hot-spot remnants'' in the Galactic halo.
 
\begin{acknowledgements} We are grateful to P. Gondolo for many helpful
discussions and thank him, C. Beck, E. Lorenz, S. Pezzoni, L.
Stodolsky, and the referees, one anonymous and especially L.O'C.Drury
for valuable comments on the manuscript.
RP is a Heisenberg
fellow of the Deutsche Forschungsgemeinschaft. 
\end{acknowledgements}

\end{document}